\documentclass{aa}
\usepackage{txfonts}
\usepackage{graphicx}
\usepackage{rotating}
\usepackage{natbib}

\begin{document}

\newcommand{\inthms}[3]{$#1^{\rm h}#2^{\rm m}#3^{\rm s}$}
\newcommand{\dechms}[4]{$#1^{\rm h}#2^{\rm m}#3\mbox{$^{\rm s}\mskip-7.6mu.\,$}#4$}
\newcommand{\intdms}[3]{$#1^{\circ}#2'#3''$}
\newcommand{\decdms}[4]{$#1^{\circ}#2'#3\mbox{$''\mskip-7.6mu.\,$}#4$}
\newcommand{\tmb}{\mbox{T$_{\rm mb}$}}
\newcommand{\OI}{\mbox{O\,{\sc i}}}
\newcommand{\dtco}{D$_{2}$CO}
\newcommand{\hdco}{HDCO}
\newcommand{\htco}{H$_{2}$CO}
\newcommand{\httco}{H$_{2}^{13}$CO}
\newcommand{\kmps}{km s$^{-1}$}
\newcommand{\ra}{\rightarrow}

\newcommand{\APer}{$\alpha$\,Per}
\newcommand{\Msun}{\,M$_{\odot}$}
\newcommand{\Mjup}{M$_{\rm Jup}$}
\newcommand{\BD}{brown dwarf}
\newcommand{\BDs}{brown dwarfs}
\newcommand{\CMD}{colour-magnitude diagram}
\newcommand{\CMDs}{colour-magnitude diagrams}
\newcommand{\CCD}{colour-colour diagram}
\newcommand{\Coll}{Collinder 359}
\newcommand{\CMC}{cluster member candidate}
\newcommand{\CMCs}{cluster member candidates}
\newcommand{\LF}{luminosity function}
\newcommand{\MF}{mass function}

\hyphenation{in-clu-ding stu-dies Fe-bru-ary Gra-na-da mo-le-cu-le mo-le-cu-les}


\title{A deep wide-field optical survey in the young open cluster Collinder 359
\thanks{Based on observations obtained at Canada-France-Hawaii Telescope}
\thanks{Table 4 is only available in electronic form
at the CDS via anonymous ftp to cdsarc.u-strasbg.fr (130.79.128.5)}
}

 \author{N. Lodieu \inst{1,2} \thanks{Visiting Astronomer, German-Spanish
                             Astronomical Centre, Calar Alto,
                             operated by the Max-Planck-Institute for 
                             Astronomy, Heidelberg, jointly with the
                             Spanish National Commission for Astronomy.}
          \and
	  J. Bouvier \inst{3}
          \and
          D. J. James \inst{3,4}
	  \and
          W. J. de Wit \inst{3}
	  \and
          F. Palla \inst{5}
	  \and
          M. J. McCaughrean \inst{2,6}
	  \and
	  J.--C. Cuillandre \inst{7}
}

   \institute{Department of Physics \& Astronomy, University of Leicester,
              University Road, Leicester LE1 7RH, UK \\
              \email{nl41@star.le.ac.uk}
            \and
              Astrophysikalisches Institut Potsdam, An der Sternwarte 16,
              14482 Potsdam, Germany
            \and
	      Laboratoire d'Astrophysique, Observatoire de Grenoble, 
              BP 53, 38041 Grenoble C\'edex 9, France  \\
              \email{Jerome.Bouvier@obs.ujf-grenoble.fr, Willem-Jan.DeWit@obs.ujf-grenoble.fr}
            \and
              Physics and Astronomy Department, Vanderbilt University, 
              1807 Station B, Nashville, TN 37235, USA \\
              \email{david.j.james@vanderbilt.edu}
            \and
              INAF, Osservatorio Astrofisico di Arcetri, Largo E. Fermi 5, 
              50125 Florence, Italy \\
              \email{palla@arcetri.astro.it}
            \and
              University of Exeter, School of Physics, Stocker Road,
              Exeter EX4 4QL, UK \\
              \email{mjm@aip.de, mjm@astro.ex.ac.uk}
            \and
              Canada-France-Hawaii Telescope Corp., 
              Kamuela, HI 96743, USA \\
              \email{jcc@cfht.hawaii.edu}
}

\offprints{N. Lodieu}

\date{Received {\today} /Accepted }

\titlerunning{A deep wide-field optical survey in Collinder 359}
\authorrunning{Lodieu et al.}

\abstract{
We present the first deep, optical, wide-field imaging survey of the
young open cluster Collinder 359, complemented by near-infrared follow-up
observations. This study is part of a large programme aimed at examining
the dependence of the mass function on environment and time.
We have surveyed 1.6 square degrees in the
cluster, in the $I$ and $z$ filters, with the CFH12K camera on
the Canada-France-Hawaii 3.6-m telescope down to completeness
and detection limits in both filters of 22.0 and 24.0 mag, respectively.
Based on their location in the optical
($I-z$,$I$) colour-magnitude diagram, we have extracted new cluster
member candidates in Collinder 359 spanning 1.3--0.03\,M$_{\odot}$,
assuming an age of 60 Myr and a distance of 450 pc for the cluster.
We have used the 2MASS database as well as our own near-infrared
photometry to examine the membership status of the optically-selected
cluster candidates.
Comparison of the location of the most massive members in Collinder 359
in a ($B-V$,$V$) diagram with theoretical isochrones suggests that
Collinder 359 is older than \APer{} but younger than the Pleiades.
We discuss the possible relationship between Collinder 359 and IC\,4665 
as both clusters harbour similar parameters, including proper motion,
distance, and age.
}
\maketitle


%
%
\section{Introduction}
\label{cr359:intro}

The number of known brown dwarfs (hereafter BDs) has increased dramatically 
over the past few years, in the field
\citep{kirkpatrick99,kirkpatrick00,burgasser02,cruz03},
as companions to low-mass stars
\citep{burgasser03a,gizis03,close03,bouy03},
in star-forming regions
\citep{lucas00,briceno02,luhman03b},
and in open clusters
\citep[][and references therein]{bouvier98,zapatero00,barrado01a,barrado02a,oliveira03}. 
Since the discovery of the first BDs in the Pleiades \citep{rebolo95,rebolo96},
numerous open clusters have been targeted in the optical and in the
near-infrared to uncover their low-mass and substellar populations,
including the Pleiades \citep{bouvier98,tej02,dobbie02a,moraux03},
$\alpha$\,Per \citep{stauffer99,barrado02a},
M\,35 \citep{barrado01a}, IC\,2391 \citep{barrado01b},
and NGC\,2547 \citep{oliveira03}.

The knowledge of the Initial Mass Function (hereafter IMF) is of prime 
importance in understanding the formation of stars. The IMF is fairly 
well constrained down to about 0.5\,M$_{\odot}$ and well approximated
by a three segment power law with
$\alpha$ = 2.7 for stars more massive than 1\Msun{},
$\alpha$ = 2.2 from 0.5 to 1.0\Msun{}, and
$\alpha$ = 0.7--1.85 in the 0.08--0.5\Msun{} mass range with a
best estimate of 1.3 \citep{kroupa02}, when expressed as the mass spectrum.
However, the mass spectrum remains somewhat uncertain in the low-mass
and substellar regimes. Most studies conducted in
Pleiades-like open clusters suggest a power law index $\alpha$ in the
range 0.5--1.0 across the hydrogen-burning limit
\citep{martin98a,bouvier98,tej02,dobbie02a,moraux03,barrado01a,barrado02a}.
A comparable result is obtained for the mass function in the solar
neighbourhood \citep{kroupa93,reid99a}.

Several theories have recently emerged to explain the formation of BDs.
First, the picture of the turbulent fragmentation of molecular clouds
can be extended to lower masses \citep{klessen01,padoan02}.
Second, \citet{whitworth04} proposed that BDs could result from the
erosion of pre-stellar cores in OB associations.
Third, gravitational instabilities of self gravitating protostellar disks
might also be responsible for the formation of BDs
\citep{watkins98a,watkins98b,lin98,boss00}.
Moreover, BDs could stop accreting gas from the molecular
cloud due to an early ejection from a multiple system 
\citep{reipurth01,bate02,delgado_donate03,sterzik03}.
Finally, as BDs straddle the realms of stars and planets,
they might form within a circumstellar disk in a similar manner
to giant planets
\citep{papaloizou01,armitage02}.

Observational evidence for disks around young BDs has been reported
in several star-forming regions in the near-infrared
\citep{muench02,wilking99,luhman99a,oliveira02},
in the L$'$-band at 3.8\,$\mu$m \citep{liu03,jayawardhana03b},
in the mid-infrared \citep{natta02,testi02,apai02},
and at millimetre wavelengths \citep{klein03}.
These results suggest a common formation mechanism for stars and BDs.
The recent lack of substellar objects in Taurus compared to the
Trapezium Cluster and IC\,348 \citep{briceno02,luhman03b}
and the distinctive binary properties of field BDs
\citep{reid01b,burgasser03a,gizis03,close03,bouy03}
hint that stars and BDs may represent two independent populations
\citep{kroupa03b}. Additional studies of nearby field BDs and
young BDs in open clusters are necessary to pin down on
the formation mechanism(s) of substellar objects.

We have initiated a large Canada-France-Hawaii Telescope Key Programme
(CFHTKP) surveying 80 square degrees in star-forming regions ($\leq$\,3 Myr),
pre-main-sequence clusters (10--50 Myr), and in the Hyades (700 Myr)
in $I$ and $z$ filters with the CFH12K wide-field camera.
The goal of our project is to investigate
in an homogeneous manner the dependence of the IMF with time and environment
as well as the distribution of stars and BDs in clusters to provide clues
on their formation. Our detection and completeness limits are
($I$,$z$)\,$\sim$\,24 and 22 mag in both passbands, respectively. 
In this paper, we present the results
of a 1.6 square degree (5 CFH12K fields-of-view) survey 
complemented by near-infrared photometry
in one young open cluster, Collinder 359 (= Melotte\,186).

Collinder 359 is a young open cluster located in the
Ophiuchus constellation around the B5 supergiant 67 Oph (HD\,164353). The cluster was
first mentioned by \citet{melotte15} as a ``large scattered group of
bright stars'' and its presence confirmed later by \citet{collinder31}.
A handful of cluster members are known \citep{collinder31,vantveer80,rucinski87}.
It lies between 200 pc (Lyng\aa{} 1987) and 650 pc
(Kharchenko 2004; personal communication) with a mean
value of 435 pc from the HIPPARCOS parallax measurement
\citep{perryman97}. The age of \Coll{} is estimated to
be about 30 Myr \citep{wielen71,abt83}.

This paper is structured as follows: A literature review of our
current knowledge of Collinder 359 is presented in \S\ref{cr359:literature}
including a discussion regarding its proper motion, distance
and age. In \S\ref{cr359:CFHTKP}, we briefly introduce the framework of the
CFHTKP. The 1.6 square degree, wide-field optical survey of
Collinder 359 is detailed in \S\ref{cr359:obs_cr359}. A description of the
candidate cluster member selection process, using an optical
($I-z$, $I$) colour-magnitude diagram, is given in \S\ref{cr359:CMCs}.
Finally, in \S\ref{cr359:IRfollowup}, we present details of a near-infrared
follow-up survey of optically-selected candidate members in Collinder 359\@.

%
%
\begin{table*}
\begin{center}
 \caption[Latest status of the bright stars in Collinder 359 (before 2000)]{
This table lists the 13 bright stars within Collinder 359 as
listed by \citet{collinder31}.
Column 1 lists the running number
of the member, column 2 gives the Henry Draper Catalogue number,
columns 3 and 4 list the right ascension and the declination
(in J2000), column 5 lists the spectral types \citep{collinder31},
columns 6, 7, and 8 lists the $V$ magnitude and the $B-V$
and $U-B$ from \citet{blanco68}, columns 9 and 10 list
the $V-R$ and $R-I$, columns 11 and 12 list
the proper motion (arcsec per year) of the object according to the SAO
catalogue (1966), column 13 gives the HIPPARCOS parallax $\pi$ \citep{perryman97}. 
The membership of the object is given
on the last column according to the discussion between
\citet{rucinski80} and \citet{vantveer80}.
}
 \begin{tabular}{ c c c c c c c c c c c c c c }
 \hline
 \hline
 \multicolumn{1}{c}{N$^{\circ}$} &
 \multicolumn{1}{c}{HD} & 
 \multicolumn{1}{c}{RA} &
 \multicolumn{1}{c}{Dec} &
 \multicolumn{1}{c}{SpT} &
 \multicolumn{1}{c}{$V$} &
 \multicolumn{1}{c}{$B-V$} &
 \multicolumn{1}{c}{$U-B$} &
 \multicolumn{1}{c}{$V-R$} &
 \multicolumn{1}{c}{$R-I$} &
 \multicolumn{1}{c}{$\mu_{\alpha}$} &
 \multicolumn{1}{c}{$\mu_{\delta}$} &
 \multicolumn{1}{c}{$\pi$} & 
 \multicolumn{1}{c}{M?} \\ \hline
 1 & 161868 & 17 47 53.5  & 02 42 26 & A0     & 3.74 &  $+$0.03 & $+$0.14  &    0.01  &    0.00 &  $-$0.0240 & $-$0.074 &   34.42$\pm$0.99 & NM  \\
 2 & 164353 & 17 58 08.3  & 02 55 57 & B5\,Ib & 3.96 &  $+$0.04 & $-$0.63  &    0.06  &    0.03 &  $-$0.0015 & $-$0.010 &    2.30$\pm$0.77 & M   \\
 3 & 164577 & 18 01 45.2  & 01 18 18 & A2     & 4.43 &  $+$0.04 & $+$0.05  &    0.04  &    0.01 &  $+$0.0090 & $-$0.012 &   12.31$\pm$0.83 & NM  \\
 4 & 164284 & 18 00 15.8  & 04 22 07 & B3     & 4.70 &  $-$0.04 & $-$0.86  &    0.10  &    0.08 &  $+$0.0000 & $-$0.013 &    4.82$\pm$0.78 & NM  \\
 5 & 166233 & 18 09 33.8  & 03 59 35 & F2     & 5.72 &  $+$0.37 & $+$0.02  &    0.22  &    0.21 &  $+$0.0360 & $-$0.007 &   19.62$\pm$1.22 & NM  \\
 6 & 165174 & 18 04 37.3  & 01 55 08 & B3     & 6.14 &  $-$0.01 & $-$0.98  &    0.03  &    0.03 &  $-$0.0045 & $-$0.003 & $-$0.76$\pm$0.89 & NM  \\
 7 & 168797 & 18 21 28.4  & 05 26 08 & B5     & 6.16 &  $-$0.02 & $-$0.64  &    0.00  &    0.01 &  $+$0.0105 & $-$0.004 &    0.97$\pm$0.83 & NM   \\
 8 & 164432 & 18 00 52.8  & 06 16 05 & B3     & 6.35 &  $-$0.08 & $-$0.77  & $-$0.01  & $-$0.01 &  $+$0.0015 & $-$0.003 &    2.17$\pm$0.87 & M   \\
 9 & 163346 & 17 55 37.5  & 02 04 29 & A3     & 6.78 &  $+$0.56 & $+$0.36  &    0.37  &    0.40 &  $-$0.0030 & $+$0.007 &    5.07$\pm$0.87 & NM   \\
10 & 164097 & 17 59 29.5  & 02 20 37 & A2     & 8.54 &  $+$0.17 & $+$0.15  &    0.12  &    0.15 &  $-$0.0060 & $+$0.003 &         ---      & M    \\
11 & 164283 & 17 57 42.4  & 05 32 37 & A0     & 9.10 &  $+$0.26 & $+$0.19  &    0.16  &    0.21 &  $+$0.0075 & $-$0.014 &    4.11$\pm$1.27 & M   \\
12 & 164352 & 18 00 41.7  & 03 08 57 & B8     & 9.33 &  $-$0.01 & $-$0.39  &    0.02  &    0.06 &  $-$0.0015 & $-$0.002 &         ---      & M   \\
13 & 164096 & 17 59 34.6  & 02 30 16 & A2     & 9.70 &  $+$0.20 & $+$0.17  &    0.13  &    0.20 &  $-$0.0105 & $-$0.006 &         ---      & M   \\
 \hline
 \end{tabular}
 \label{tab_cr359:old_members}
\end{center}
\end{table*}
%

%
%
\section{The open cluster Collinder 359}
\label{cr359:literature}

Collinder 359 was selected as a 30 Myr
pre-main-sequence open cluster at a distance of 250 pc within
the framework of our CFHTKP using using data obtained from the {\sc{webda}}
open clusters database\footnote{The {\sc webda} database, maintained
by J.-C.\ Mermilliod, is to be found at http://obswww.unige.ch/webda/}.
The equatorial and Galactic coordinates
(J2000) of the cluster centre are
(18$^{\rm h}$01$^{\rm m}$,$+$02$^{\circ}$54$'$) and (29.7,$+$12.5),
respectively. Very little is known about \Coll{}
and no deep optical survey has been conducted in the cluster to date.
We give below a brief overview of the current knowledge on
the cluster.

Collinder 359 was first seen on the Franklin-Adams Charts Plates
by \citet{melotte15} and listed in his large catalogue
of globular and open clusters.
\citet{melotte15} classified \Coll{} as a ``coarse cluster'' 
and described it
as ``a large scattered group of bright stars around 67 Ophiuchi,
covering an area of about 6 square degrees''.
In his catalogue of open clusters, \citet{collinder31}
described it as ``a group of about 15 stars with no appreciable
concentration on the sky and no well-defined outline'' and added that
``cluster stars appear brighter than the surrounding stars but no
bright stars stand out from the others''. The diameter of the cluster
was estimated to 240 arcmin and dimensions  
of 5$^{\circ}$\,$\times$\,3$^{\circ}$ were also mentioned.

\citet{collinder31} listed 13 cluster members and provided
coordinates, photometry, spectral types, and
proper motion information when available
(Table \ref{tab_cr359:old_members};
filled circles in Fig.\ \ref{fig_cr359:fc_coll359}). 
Additional photometry was compiled
by \citet{blanco68}. Isochrone fitting to five early-B stars
yielded photometric parallax of 0.0048 (d = 209 pc) while the fainter B8--A2
stars gave a mean parallax of 0.0035 (d = 286 pc).
However, the membership of these objects is not well established
and triggered a discussion based on a new CCD photometry between 
\citet{rucinski80}, \citet{vantveer80}, and \citet{rucinski87}.
Later, \citet{baumgardt00} combined photometry, radial velocities,
and parallaxes to reject some objects originally proposed as members.

\subsection{The cluster proper motion}
\label{cr359:literature_motion}

\Coll{} has a small proper motion according to the original study by
\cite{collinder31} and the more recent measurement from the astrometric
satellite HIPPARCOS\@. The proper motion of the cluster is comparable to
the motion of 67 Oph, its most massive member, and was estimated
as 0.42\,$\pm$\,0.47 mas/yr in right ascension and
$-$7.86\,$\pm$\,0.35 mas/yr in declination \citep{baumgardt00}.
Similarly, \citet{perryman97} quotes a proper motion of
(0.41,$-$8.22) mas/yr.

\begin{figure}[!h]
\begin{center}
\includegraphics[width=\linewidth, angle=0]{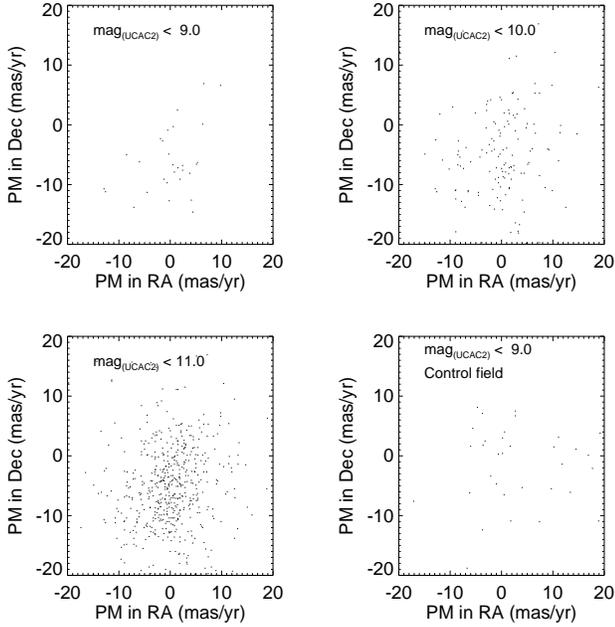}
\bigskip
\caption[Vector point diagrams for Collinder 359]{
Vector point diagrams for all stars within two
degrees in radius from the cluster centre for magnitudes
brighter than 9, 10, and 11, respectively. The bottom
right displays the vector point diagram from a
control field
(RA = 17$^{\rm h}$08, dec = $+$9$^{\circ}$).
The proper motions (accurate to 6 mas/yr)
are taken from the USNO CCD Astrograph Catalog
(\citealt{zacharias04}). Two clusterings are clearly
separated for magnitudes brighter than 10.0\@.
The first is located at (0,0), and the second at approximately
(0.0,$-$8.5) mas/yr in right ascension and declination, respectively.
The comparison between the cluster and controls fields at
bright magnitudes depicts the presence of an overdensity
of sources with a proper motion consistent with \Coll{}.
}
\label{fig_cr359:coll_PM_evol}
\end{center}
\end{figure}

\begin{figure}[!h]
\begin{center}
\includegraphics[width=\linewidth, angle=0]{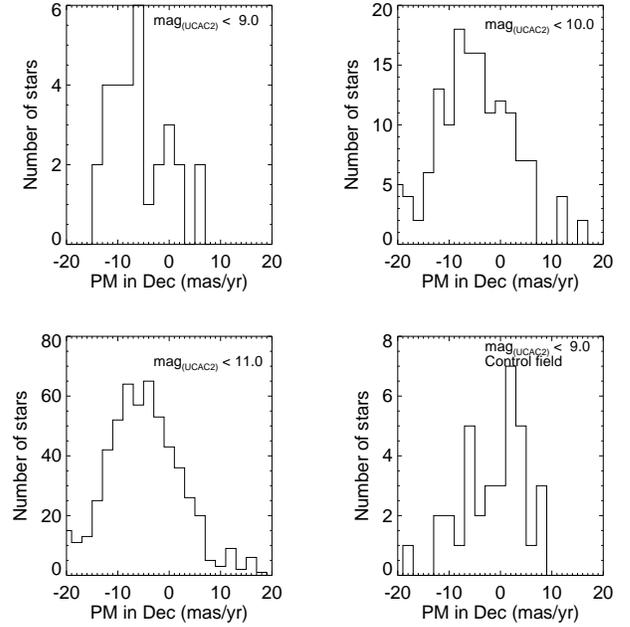}
\bigskip
\caption[]{Histograms of the number of sources as a function of
declination for all stars located within two
degrees in radius from the cluster centre for magnitudes
brighter than 9, 10, and 11, respectively.
We have added a vector point diagram for a control field
over a 2 degree radius centred on RA = 17$^{\rm h}$08
and dec = $+$9$^{\circ}$ (bottom right).
The proper motions are from the UCAC catalogue
(\citealt{zacharias04}) and accurate to 6 mas/yr. 
The presence of the cluster is inferred from the
comparison of the top left and bottom right histograms.
The top right histogram shows an overdensity of sources
at $\mu_\delta \sim -$8.0 mas/yr.
}
\label{fig_cr359:coll_PM_evol_hist}
\end{center}
\end{figure}

We have used the second release of the USNO CCD Astrograph Catalog
\citep[UCAC2;][]{zacharias04} project to illustrate the cluster 
proper motion.
Figure \ref{fig_cr359:coll_PM_evol} displays the vector point diagrams
(proper motion in right ascension versus proper motion in declination)
for all stars within two degrees in radius from the cluster centre
for magnitude brighter than 9.0, 10.0, and 11.0, respectively.
Two clustering of stars emerge from these diagrams.
Similarly, two peaks are also seen when plotting the histogram 
of the number of stars as a function of declination
for magnitudes brighter than 10.0 mag.
(Fig.\ \ref{fig_cr359:coll_PM_evol_hist}).
The first group of stars has no significant proper motion
and denotes field stars whereas the second exhibits a shift
in declination and corresponds to the cluster.
In addition, we have included the vector point diagram for
a control field (RA = 17$^{\rm h}$08, dec = $+$9$^{\circ}$)
over a 2 degrees in radius and
for UCAC2 magnitudes brighter than 9.0\@. The comparison
of this diagram with the top left one in 
Fig.\ \ref{fig_cr359:coll_PM_evol}
reveals an overdensity of stars at expected proper motion
for \Coll{}. The top right histogram in 
Fig.\ \ref{fig_cr359:coll_PM_evol_hist} shows that the number
of sources at $\mu_\delta \sim -$8.0 mas/yr exceeds the number
of objects with a zero proper motion, confirming hence the
presence of a cluster.

%
%
%

%
%
\subsection{The age of the cluster}
\label{cr359:literature_age}

\cite{wielen71} derived an age of 20--50 Myr with
a mean value of 30 Myr by fitting isochrones in three-colour
photometry obtained from large catalogues of open clusters \citep{becker71}.
\citet{abt83} studied the distribution of Ap stars in open clusters
as a function of age and put an upper limit
of 30 Myr on the age of \Coll{}, assuming that 67 Oph is
a member of the cluster. Both results are in agreement and consistent
with the most recent estimate from Kharchenko et al.\ (2004; personal communication).

\begin{figure}[!h]
\begin{center}
\includegraphics[width=\linewidth, angle=0]{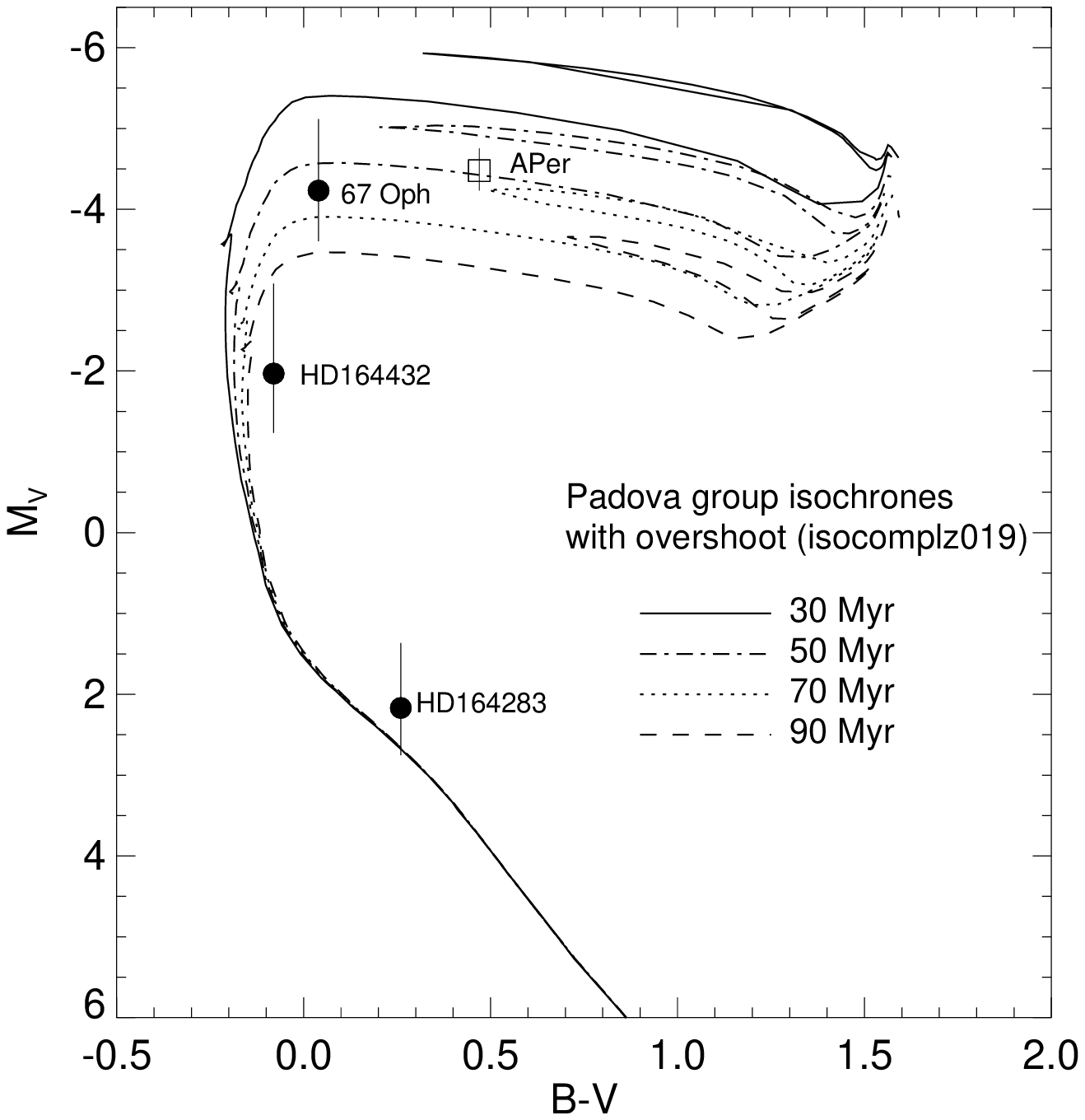}
\caption{
($B-V$,M$_{V}$) \CMD{}. The position of the F5 supergiant
Alpha Per (open square), the B5 supergiant 67 Oph, 
HD164432 (B3), and HD164283 (A0)
(filled circles) are indicated. Overplotted are the
solar metallicity evolutionary models with moderate overshoot
from the Padova group \citep{girardi00} for
30 Myr (solid line), 50 Myr (dot-dashed line),
70 Myr (dotted line), and 90 Myr (dashed line).
The vertical line crossing the solid circles represent the
errors on the HIPPARCOS parallax for 67 Oph
\citep{perryman97}.
The best fit is obtained for ages of 50 Myr and 60 Myr for
the \APer{} and \Coll{} clusters, respectively.
}
\label{fig_cr359:coll_girardi}
\end{center}
\end{figure}

We have attempted to estimate the age of \Coll{} and its associated error
using its most massive members. We have followed the approach
applied to \APer{} by \citet{stauffer03}.
By comparing the location of the star $\alpha$\,Persei
(open square in Fig.\ \ref{fig_cr359:coll_girardi}) in the \CMD{}
($B-V$,M$_{V}$) with theoretical
solar metallicity isochrones including moderate overshoot
(\citealt{girardi00}), an age of 50 Myr was inferred for \APer{}
(dot-dashed line in Fig.\ \ref{fig_cr359:coll_girardi}).
We should keep in mind that the lithium test applied to the
\APer{} cluster yielded a value larger than the turn-off
main-sequence method (90 Myr versus 50 Myr; \citealt{stauffer99}).

%
%

\Coll{} is located around the B5 supergiant, 67 Oph
(filled circle in Fig.\ \ref{fig_cr359:coll_girardi}), which
is considered as a member of the cluster with a probability of
75\,\% and over 95\,\% by \citet{baumgardt00}
and Kharchenko et al.\ (2004), respectively.

Assuming a mean apparent magnitude of $V$ = 3.96\,$\pm$\,0.02
and a mean distance of 435$^{+220}_{-110}$ pc \citep{perryman97},
we have derived an absolute magnitude of
M$_{V}$ = $-$4.23$^{+0.63}_{-0.89}$. The vertical line crossing
the circles in Fig.\ \ref{fig_cr359:coll_girardi} represents the uncertainty
on the parallax estimate of 67 Oph. The best positional fit
of 67 Oph in the ($B-V$,M$_{V}$) \CMD{} is obtained for
an age of 60 Myr (between the dot-dashed and dotted lines in
Fig.\ \ref{fig_cr359:coll_girardi}),
with an uncertainty of 20 Myr (extent of the vertical line).
Our age estimate is twice the value of 30 Myr quoted by
\citet{wielen71} but larger than the main-sequence turn-off
age of the \APer{} cluster, suggesting that \Coll{} is likely
older than \APer{} but younger than the Pleiades.
We note that more recent age determination in open clusters
using the lithium test \citep*{rebolo92} led to older ages
by a factor of $\sim$\,1.6 than the turn-off main-sequence
method \citep{jeffries01a}. The age of Collinder 359 could 
therefore be as old as 100 Myr.

\subsection{The distance of the cluster}
\label{cr359:literature_distance}

Based on isochrone fitting of early-type stars, \cite{collinder31} derived
a distance ranging from 210 pc to 290 pc for \Coll{}.
Following new photometric studies by \citet{vantveer80} and 
\citet{rucinski80,rucinski87}, the rejection of some objects originally
proposed as cluster members yielded a revised distance of 436 pc.
In the 5$^{\rm th}$ Open Cluster Data Catalogue (Lyng\aa{} 1987),
a distance of 200 pc was quoted based on the Bochum-Strasbourg
magnetic tape catalogue of open clusters. The HIPPARCOS parallax
measurement of the supergiant 67 Oph
led to a distance of 435$^{+220}_{-110}$ pc
\citep{perryman97}. Trigonometric parallaxes of five photometric
members from HIPPARCOS yielded distances between 260 and 280 pc
for \Coll{} \citep{loktin01}.
Combining the HIPPARCOS and Tycho\,2 catalogues, a list
of about 100 possible cluster members (open squares in
Fig.\ \ref{fig_cr359:fc_coll359}) were extracted by
Kharchenko et al.\ (2004, personal communication) based on their
location within the cluster area and their proper motions.
The position of these objects in the ($B-V$,$V$) \CMD{}
yielded a distance of 650 pc from isochrone fitting.

To summarise, the distance of \Coll{} is not well constrained to date.
We will adopt a mean distance of 450 pc with an uncertainty
of 200 pc (Table \ref{tab_cr359:dist}).

%
%
\begin{table}[!h]
\begin{center}
 \caption{
Method and references for the various distance estimates
for the open cluster \Coll{}. In this paper, we adopt a mean 
distance of 450 pc with an uncertainty of 200 pc.
}
 \begin{tabular}{@{}l c l@{}}
 \hline
 \hline
Method &  d (pc) & reference \\
 \hline
Isochrone fitting    & 210--290 & \cite{collinder31} \\
Isochrone fitting    & 436      & \citet{rucinski87} \\
Isochrone fitting    & 200      & Lyng\aa{} 1987     \\
Parallax (HIPPARCOS) & 435$^{+220}_{-110}$  & \cite{perryman97} \\
Parallax (HIPPARCOS) & 260--280 & \cite{loktin01}    \\
Isochrone $+$ proper motion     & 650       & Kharchenko et al.\ 2004 \\
 \hline
 \end{tabular}
 \label{tab_cr359:dist}
\end{center}
\end{table}
%

%
%
\section{The CFHT Key Programme}
\label{cr359:CFHTKP}

We have carried out a CFHT Key Programme
(30 nights over 2 years) centred on wide-field optical imaging
in a variety of environments to examine the sensitivity
of the low-mass stellar and substellar IMF to time
and environment. This work was conducted 
within the framework of the European Research Training Network
``The Formation and Evolution of Young Stellar Clusters''.
Other goals were to address the most pressing
issues concerning low-mass stars and BDs, including their formation,
their distribution, and their evolution with time.
The survey was conducted with a large-CCD mosaic
camera (CFH12K) in the $I$ and $z$ filters down to
detection and completeness limits of $I$ = 24.0 and
22.0 mag, respectively, covering a total of 80 square degrees
in star-forming regions, open clusters, and in the Hyades.

The CFH12K is a CCD mosaic camera dedicated to
high-resolution wide-field imaging.
The camera comprises 12 chips of 4128\,$\times$\,2080 pixels with a
pixel scale of 0.206 arcsec, yielding a field-of-view
of 42\,$\times$\,28 arcmin. Hence,
no problem of undersampling was foreseen even during excellent
conditions on Mauna Kea, which was the case for our observations.
The cosmetics of the CFH12K mosaic was excellent with
a total of 200 bad columns, most of them were concentrated
on CCD05. The CCD06, CCD08, CCD09, CCD10, and CCD11 are entirely
free of bad columns.

We have chosen to carry out the wide-field
optical observations in the $I$ and $z$
filters mainly to optimise the search for low-mass
stars and BDs in young clusters. This choice was
also motivated by the results of a 6.4 square degree imaging survey
of the Pleiades with the CFH12K in the $I$- and $z$-bands
\citep{moraux03} conducted with the same telescope/instrument
configuration. Numerous brown dwarf candidates in the Pleiades were discovered
down to 30\,\Mjup{}. The derived mass function was in agreement
by that determined by \cite{bouvier98}. Moreover, the empirical
knowldge of the mass function was extended far into the substellar
regime.

%
%
%
\section{The wide-field optical survey in Collinder 359}
\label{cr359:obs_cr359}
%

%
%
\subsection{Observations and data reduction}
\label{cr359:obs_obs}

\begin{figure}[htbp]
\includegraphics[width=\linewidth, angle=0]{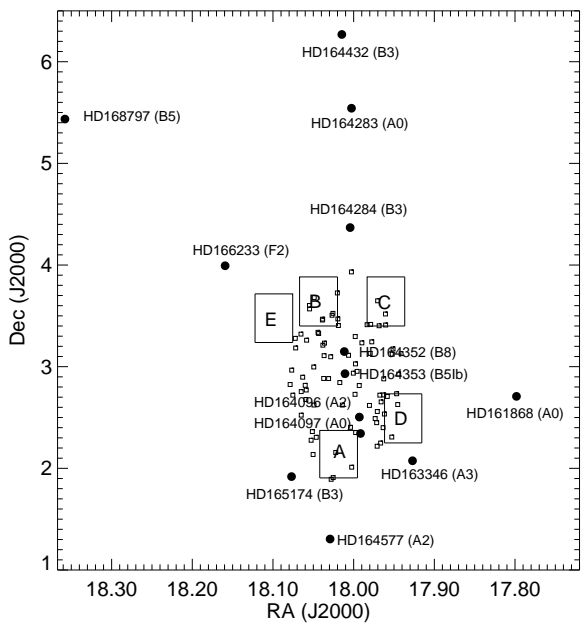}
\caption{
Location of the five CFH12K fields-of-view (A, B, C, D, and E)
shown as boxes within the cluster area defined by the {\sc{webda}} Database.
The 13 possible cluster members listed by \citet{collinder31}
and given in Table \ref{tab_cr359:old_members}
are displayed as filled circles. Their names and spectral types are
provided as well. The open squares are possible members
used for isochrone fitting by Kharchenko et al.\ (2004; personal communication)
to derive a distance of 650 pc and an age of 30 Myr.
}
\label{fig_cr359:fc_coll359}
\end{figure}

Five CFH12K frames were obtained on 18 
and 20 June 2002 in \Coll{} in the $I$ and $z$ filters, 
covering a total area of 1.6 square degrees in the cluster
(Table \ref{tab_cr359:log_obs}).
Figure \ref{fig_cr359:fc_coll359} displays the location of the five CFH12K
pointings within the cluster area and Table \ref{tab_cr359:log_obs}
gives the journal of the observations. Thirteen possible members as
listed by \citet{collinder31} (filled circles in Fig.\ \ref{fig_cr359:fc_coll359})
are also overplotted.
The CFH12K frames were chosen to avoid bright cluster members.
These frames are located about half a degree away from the cluster
centre and have a small overlap with the possible members extracted
by Kharchenko et al.\ (2004; personal communication).

%
%
\begin{table}[!h]
\begin{center}
 \caption{
Coordinates (J2000) of the five CFH12K fields-of-view 
(42 arcmin $\times$ 28 arcmin) along with
the journal of observations obtained in the pre-main-sequence open cluster \Coll{}.
The times of observations are given in UT and correspond to
the beginning of the short exposures in the $I$-band.
}
 \begin{tabular}{c c c c c c c }
 \hline
 \hline
Field &  R.A.\     & Dec         & Obs.\ Date  & Time of obs.\ \cr
 \hline
  A   & 18 01 06.6 & $+$02 07 26 & 2002--06--18 & 08h19m15s \cr
  B   & 18 02 36.9 & $+$03 37 52 & 2002--06--18 & 09h07m43s \cr
  C   & 17 57 36.9 & $+$03 37 56 & 2002--06--18 & 09h56m07s \cr
  D   & 17 56 16.4 & $+$02 29 46 & 2002--06--18 & 11h52m24s \cr
  E   & 18 05 55.7 & $+$03 28 58 & 2002--06--20 & 12h29m20s \cr
 \hline
 \end{tabular}
 \label{tab_cr359:log_obs}
\end{center}
\end{table}

Fields A, B, C, and D were obtained
on 18 June 2002 under photometric conditions with seeing
$\sim$\,0.8 arcsec. The remaining field, field E, was observed
on 20 June 2002 under non-photometric conditions
(extinction less than 0.05 mag at the time of the observations;
see atmospheric attenuation on the Elixir 
webpage\footnote{http://www.cfht.hawaii.edu/Instruments/Elixir/stds.2003.06.html}).
Three sets of exposures were taken for
each field-of-view: short, medium, and long exposures with
integration times of 2, 30, and about 900 seconds, respectively.
The long exposures consisted of three times 300 and 360 seconds in the
$I$ and $z$ filter, respectively, yielding detection limits
of 24.0 mag in both passbands.
Only one image was taken for the short and medium exposures,
whereas three dithered positions were obtained for the long
exposures, allowing rejection of bad pixels and removal
of bad columns.
The observations were scheduled in a queue mode so that the short,
medium, and long exposures in the $I$-band were taken
immediately prior to the short, medium, and long exposures
in the $z$-band.

%
%

The initial data reduction was provided by the Elixir pipeline 
\citep*{magnier04} and
includes bias subtraction, flat-fielding, correction for
scattered light in the $I$ and $z$ bands, combining
the dithered frames in case of long exposures, and astrometric solution
provided in the header of the fits files. Standard stars
were observed throughout the nights and were monitored constantly
by the Elixir/Skyprobe tool to provide accurate zero-points.

%
%
\subsection{Optical photometry}
\label{cr359:obs_phot}

We have used the SExtractor
software\footnote{http://astroa.physics.metu.edu.tr/MANUALS/sextractor/}
\citep{bertin96} to extract the photometry from the optical images.
We have favoured the point-spread function (PSF) fitting to aperture
photometry because it provides more precise photometric measurements
for faint sources, which are, in our case, the cluster \BD{} candidates.
We have been kindly provided the PSFex package by
E.\ Bertin (personal communication) to carry out PSF fitting.

The data reduction procedure to extract a catalogue of all objects
from the reduced and stacked images processed by the pipeline was
identical for each CFH12K field-of-view. First, we have combined the $I$
and $z$ images to increase the signal-to-noise ratio and permit
a better astrometry of faint sources close to the detection
limit. Therefore, the detection of sources was run on the combined
image whereas the photometry was applied to the $I$- and $z$-band images.
We have run SExtractor and PSFex
to extract coordinates and magnitudes for all detected sources
by using a model PSF for each chip and each field.
Afterwards, we have applied the zero points listed on the Elixir
webpage\footnote{http://www.cfht.hawaii.edu/Instruments/Elixir/stds.2003.06.html}
to calibrate our photometric measurements. The nominal CFH12K
zero points for the $I$- and $z$-bands were
ZP($I$) = 26.184\,$\pm$\,0.023 and ZP($z$) = 25.329\,$\pm$\,0.031,
respectively. Small corrections were applied for the nights of 
18 and 20 June 2002 to take into account the weather conditions on
those specific nights.
Finally, the $I$ and $z$ catalogues were cross-correlated
by matching pixel coordinates.

One catalogue was generated for each CCD chip of each CFH12K
field-of-view for all three exposures (short, medium, and long).
The catalogues contain the pixel and celestial coordinates,
magnitudes, as well as other
parameters, including the full-width-half-maximum and the 
ellipticity of the source. The final magnitudes (see electronic
table) have rms errors on the $I$-band magnitudes smaller than 0.1 mag.

%
%
\subsection{Calibration of the photometry}
\label{cr359:obs_phot_calib}

We have conducted a number of ``checks'' to verify the validity of the photometry.
We have ensured that colour-magnitude diagrams of all chips in one 
CFH12K field-of-view align properly \citep[see][]{deWit05}.
To calibrate internally our photometry, we have cross-correlated 
the short and medium and the medium and long exposures for each individual 
field-of-view. We were unable to calibrate the magnitudes between fields 
because no overlapping area between the CFH12K pointings was available.

%
%
\begin{figure}[!h]
\begin{center}
\includegraphics[angle=0, width=\linewidth]{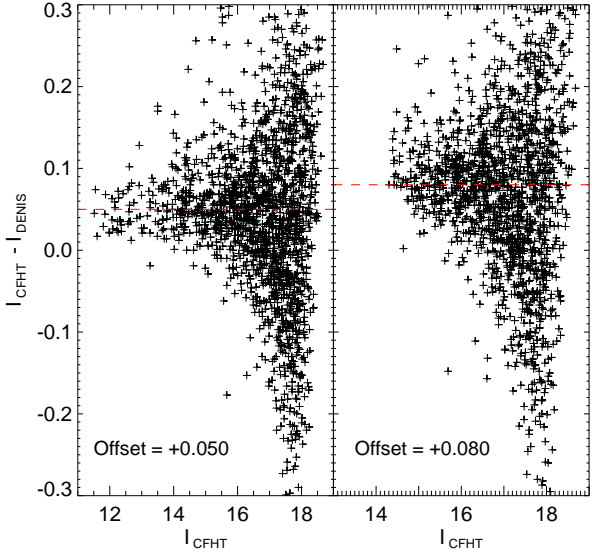}
\caption[Offsets between the CFHT and DENIS $I$ magnitudes]{
Offsets in $I$ magnitudes between the CFHT and DENIS measurements
for a 15\,$\times$\,6 arcmin overlapping area
located in field A\@. Photometric shifts of $+$0.050 and $+$0.080 mag
are found for the short (left panel) and medium (right panel)
exposures, respectively. The same procedure was not possible
with the deep exposures since the detection of DENIS corresponds
to the saturation limit of the CFH12K long exposures.
}
\label{fig_cr359:offsets_CFHT_DENIS}
\end{center}
\end{figure}
%

%
%
%
%
%
\begin{table*}[t]
\begin{center}
 \caption[]{ 
Catalogue of cluster member candidates in \Coll{}, assuming a distance
of 450 pc and an age of 60 Myr for the cluster. The total number of
sources is 506\@. The name of each individual target is given in
the first column according to the IAU nomenclature.
The CFH12K field-of-view and the identification number of each
source is provided in Cols.\ 2 and 3\@. Right ascension and
declination (in J2000) are listed in Cols.\ 4 and 5\@.
Optical ($I$ and $z$) and 2MASS near-infrared magnitudes are
provided in Cols.\ 6--10\@. Finally, the membership of each
target is given in the last column. Y stands for possible member
whereas NM depicts a non-member of the cluster.
}
 \begin{tabular}{c c c c c c c c c c c}
 \hline
 \hline
IAU Name  &  FOV  & ID  & R.A.\  &  Dec &  $I$ & $z$ & $J$ & $H$ & $K_{s}$ & Mem? \cr
 \hline
Cr359~J175453$+$023353 & D00 &  1469 & 17 54 53.20 & 02 33 53.6 & 11.537 & 11.326 & 10.197 &  9.457 &  9.236 & Y \\
\ldots & \ldots & \ldots & \ldots & \ldots & \ldots & \ldots & \ldots & \ldots & \ldots & \ldots \cr
Cr359~J175507$+$022403 & D06 &  5613 & 17 55 07.79 & 02 24 03.2 & 22.480 & 21.320 &  ---   &  ---   &  ---   & Y \\
 \hline
 \end{tabular}
 \label{tab_cr359:coll_all_CMCs}
\end{center}
\end{table*}

We have also attempted to calibrate our photometry externally.
However, on the one hand, the CFH12K
fields-of-view were chosen to avoid bright cluster members
(Fig.\ \ref{fig_cr359:fc_coll359}) and, on the other hand,
no previous study was conducted in those areas.
Nevertheless, we have cross-correlated our
final source catalogue with the recent release of the
DEep Near-Infrared Survey \citep{epchtein97}.
We could extract a small overlapping region (15\,$\times$\,6 arcmin)
between the DENIS survey and the area covered by Field A
(17$^{\rm h}$59$^{\rm m}$30$^{\rm s}$\,$\leq$\,RA\,$\leq$\,18$^{\rm h}$02$^{\rm m}$30$^{\rm s}$
and $+$01$^{\circ}$54\,$\leq$\,Dec\,$\leq$\,$+$02$^{\circ}$00).
The mean photometric offsets in the $I$ magnitudes between the CFHT
and DENIS measurements are $+$0.05$\pm$0.09 and $+$0.08$\pm$0.12 mag
for short and medium exposures, respectively,
as shown in Fig.\ \ref{fig_cr359:offsets_CFHT_DENIS}.
A similar procedure could not be applied to the long exposures
because the DENIS detection limit corresponds to the
saturation of the CFH12K long exposures ($I$ = 18.0--18.5 mag).
The agreement between the DENIS and CFH12K magnitudes is
on the order of the dispersion observed in our photometry
and does not affect the subsequent candidate selection in \Coll{}.
Note that the errors on the DENIS magnitudes are up to 0.13, 0.18,
and 0.22 mag at $I$ = 16, 17, and 18, respectively.

%
%
\section{New cluster member candidates in \Coll{}}
\label{cr359:CMCs}
\subsection{Optical colour-magnitude diagram}
\label{cr359:CMCs_CMD}

\begin{figure*}
\begin{center}
\includegraphics[angle=0, width=0.74\linewidth]{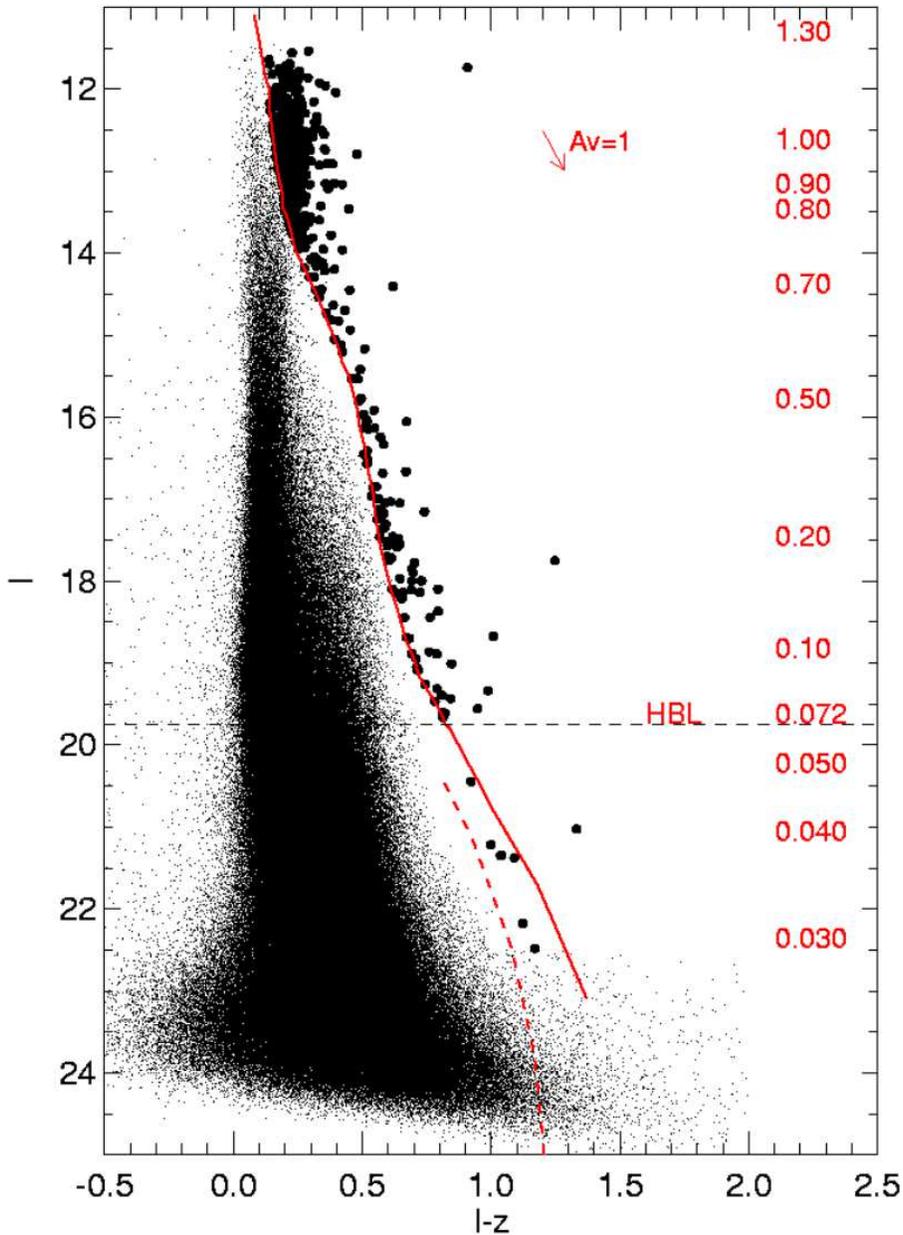}
\caption{
Colour-magnitude diagram ($I-z$,$I$)
for the intermediate-age open cluster \Coll{} over the full
1.6 square degree area surveyed by the CFH12K camera.
The large filled dots are all optically-selected
cluster member candidates spanning 1.3--0.03\Msun{},
assuming an age of 60 Myr and a distance of 450 pc.
Overplotted are the NextGen \citep[solid line;][]{baraffe98}
and the DUSTY \citep[dashed line;][]{chabrier00c}
isochrones for 60 Myr, assuming a distance of 450 pc for the
cluster. The dashed line at $I$\,$\sim$\,19.75 mag indicates
the hydrogen-burning limit (HBL) at 0.072\Msun{}.
The mass scale (in \Msun{}) is given on the right side of the graph
for the assumed age and distance for the cluster.
A reddening vector of A$_{V}$ = 1 mag is indicated for
comparison purposes.
}
\label{fig_cr359:coll_cmd_total}
\end{center}
\end{figure*}

The final ($I-z$,$I$) \CMD{} for all detections in the 1.6
square degree area surveyed in \Coll{} is presented in
Fig.\ \ref{fig_cr359:coll_cmd_total}.
The detection and completeness limits of the survey
are estimated to $I$\,$\sim$\,$z$\,$\sim$\,24.0 and 22.0 mag,
respectively. To create the final \CMD{}, we have cross-correlated
the short with medium and medium with long exposures to
remove common detections and keep the best photometry.
Hence, the photometry of the objects with $I$\,$\leq$\,15.0 mag,
15.0\,$\leq$\,$I$\,$\leq$\,19.0 mag,
and $I$\,$\geq$\,19.0 mag is extracted from the short, medium, and
long exposures, respectively.

%
%

Overplotted on the \CMD{} are the 60 Myr NextGen \citep[solid line;][]{baraffe98}
and the DUSTY \citep[dashed line;][]{chabrier00c}
isochrones, assuming a distance of 450 pc for the cluster.
The horizontal dashed line at $I$\,$\sim$\,19.75 mag corresponds to
the hydrogen-burning limit (HBL) at 0.072\Msun{}.
The mass scale indicated on the right-hand side of the
plot in solar masses spans 1.3\Msun{} to 0.030\Msun{}.
A reddening vector with A$_{V}$ = 1 mag is indicated by an arrow
for comparison purposes. We have considered an interstellar absorption
law with A$_{I}$ = 0.482 mag for the $I$-band \citep*{rieke85}.
As no estimate is available in the $z$ band, we have assumed a linear
fit between the interstellar absorption in the $I$ and $J$ bands
(A$_{J}$ = 0.282 mag), yielding A$_{z}$ = 0.382 mag.

\subsection{Selection of cluster member candidates}
\label{cr359:CMCs_select}

The extraction of member candidates in open clusters
generally consists in selecting objects located to the right of
the ZAMS (\citealt{leggett92}) shifted to the distance of the cluster.
We have chosen the evolutionary models from the Lyon group to select
candidates in \Coll{}. We have employed the NextGen isochrones
(solid line in 
Fig.\ \ref{fig_cr359:coll_cmd_total};~\citeauthor{baraffe98}~\citeyear{baraffe98})
for effective temperatures higher than 2500\,K (corresponding to
0.050\Msun{} at the age and distance of the cluster) and the DUSTY
(dashed line in 
Fig.\ \ref{fig_cr359:coll_cmd_total};~\citeauthor{chabrier00c}~\citeyear{chabrier00c})
isochrones for lower temperatures (and masses). We did not consider the Cond models
since the isochrone lie to the right of the DUSTY isochrone.
Consequently, objects located to the right of the Cond isochrones
are to the right of the DUSTY isochrones as well and, hence, remain
bona-fide \CMCs{}.
We have selected {\it{all}} objects located to the right of the combined
NextGen$+$DUSTY isochrones, assuming a distance of 450 pc and an
age of 60 Myr for \Coll{}.

We have examined each cluster candidate by eye both in the $I$ and $z$
images to reject extended objects, blended sources, and detections
affected by bad pixels or bad columns. Indeed, more than two-thirds of
the objects located to the right of the evolutionary models
were affected by bad pixels in one filter at least, or located
on a bad column despite the good cosmetics of the CFH12K
camera. 

After removal of all spurious detections, the final
list of cluster members contains a total of 506
candidates ranging from $I$ = 11.3 to $I$ = 22.5 mag over
1.6 square degree area surveyed in \Coll{}
(Table \ref{tab_cr359:coll_all_CMCs}).
Column\,1 gives the name of the target according to the IAU
convention. We used the name Coll359\,J to refer to \Coll{} followed
by the coordinate in J2000\@.
Columns 2 and 3 provide the field-of-view where the candidate
is located and the identification number assigned during
the extraction of the photometry.
Cols.\ 4 and 5 list the right ascension (in hours) and declination
(in degrees) of the objects extracted from the CFH12K images (in J2000).
Cols.\ 6--11 give the optical ($I$ and $z$) and the
near-infrared ($J$, $H$, and $K_{s}$) magnitudes.
Column 12 provides an update of the membership status
of the candidate after considering the near-infrared follow-up,
where Y stands for possible members and NM for non-members.

The range of ellipticities and full-width-half-maxima for all candidates
are 0.001--0.265 and 1.6--3.2, respectively.
Only one object has an ellipticity of 0.602 and a FWHM of 2.8,
casting doubt about its membership.
The distribution of ellipticities shows that 95\,\% of the objects have
an ellipticity smaller than 0.15\@. The majority of objects have
FWHM between 1.8 and 3.0, corresponding to a seeing of
$\sim$ 0.4--0.6 arcsec.

It is possible that we have missed some bona-fide cluster
members for various reasons. Our detection rate was not 100\,\% 
nor was our completeness limit sufficiently deep to have detected
all members. Additionally, we did not detect the bright cluster
members due to our saturation limit at about $I$ = 12 mag.
Second, 200 bad columns affect the CFH12K field-of-view and
most especially the CCD05 where the largest incompleteness is 
expected. Similarly, objects affected by bad pixels
might actually be genuine cluster candidates but were rejected 
from the final list because of their dubious photometry.
Source blending represents another effect which prevents us
from detecting all members. Last but not least, we did not cover the 
whole cluster, implying that there are other, potentially large
numbers of members yet to be discovered.

From the colour-magnitude diagram ($I-z$,$I$), the large
field contamination at magnitudes brighter than
$I$\,$\sim$\,14 mag is clearly visible. Out of the
506 candidates, 70\,\% of them lie in the range $I$ = 12--14 mag.
The large number of candidates at brighter magnitudes (and thus at high masses) 
originates from the merging between the cluster sequence and the
sequence of field stars. There is likely to be considerable field
star contamination in this mass range for the cluster as seen
in the colour-magnitude diagram.

If we consider an age of 100 Myr as suggested by recent age determination 
in open clusters using the lithium test \citep{stauffer99},
the number of candidates increases to 628 in the same magnitude range.
According to evolutionary isochrones \citep{baraffe98}, the stellar/substellar
boundary would then be at $I$ = 20.2 mag, half a magnitude fainter
than at 60 Myr.

%
%
\section{Near-infrared photometry follow-up of optically-selected candidates}
\label{cr359:IRfollowup}

Collinder 359 is at a Galactic latitude
of $|$b$|$ = 12.5$^{\circ}$, intermediate between \APer{}
($|$b$|$ = 7$^{\circ}$) and the Pleiades ($|$b$|$ = 24$^{\circ}$).
Therefore, the sample of optically-selected \CMCs{} in \Coll{}
is inevitably contaminated by foreground and background objects,
including galaxies, reddened background giants, and field dwarfs.

We took special care in the removal of extended objects
from the cluster candidate list so that we expect a small
contamination by background galaxies.
Field dwarfs also represent
another source of contamination as they have similar optical 
colours as young 
cluster members. However, optical-to-infrared \CMDs{} such
as ($I-J$,$I$) or ($I-K$,$I$) have proven their efficiency
in weeding out field dwarfs in $\sigma$\,Orionis \citep{zapatero00},
in the Pleiades \citep{zapatero97b,pinfield00}
in \APer{} \citep{barrado02a}, and in IC\,2391 \citep{barrado01b}.
Furthermore, the latest theoretical DUSTY isochrones \citep{chabrier00b}
predict bluer $I-K$ colours for field dwarfs than young
low-mass cluster members by 1.0 to 1.5 mag depending on the mass.

\subsection{Cross-correlation with the 2MASS database}
\label{cr359:IRfollowup_2MASS}

To estimate the contamination towards
bright member candidates in \Coll{}, we have cross-correlated
the sample of optically-selected \CMC{} with the 2MASS
All-Sky release catalogue of point-sources \citep{cutri03}
\footnote{http://www.ipac.caltech.edu/2mass/releases/second/doc/}.
Due to its completeness limit of $K_s$ = 14.3 mag,
the 2MASS database provides infrared counterparts in
$J$, $H$, and $K_{s}$ for most of the optically-selected
cluster candidates brighter than $I$ = 17.0 mag.
For objects fainter than $K_s$ = 14.3 mag, the uncertainty on
the magnitude become larger than 0.1 mag
and additional near-infrared observations are required
to establish membership.

Among 433 \CMCs{} brighter than 17.0 mag in the $I$-band, 426 of them
have a 2MASS counterpart within a radius of 2 arcsec, with
$K_s$\,$<$\,14.3, and errors on the
$J$, $H$, and $K_s$ magnitudes smaller than 0.1 mag.
Robust near-infrared photometry is available for 97\,\% of our
sample to $I$ = 17 mag.
The remaining objects have a 2MASS counterpart
but either at larger radii (2 to 3 arcsec) or $K_s$ magnitudes
fainter than 14.3 or uncertainties larger than 0.1 mag.
We have therefore not considered their 2MASS
magnitudes but obtained additional photometry for half of them.

\subsection{Additional near-infrared photometry}
\label{cr359:IRfollowup_CFHT}

We have carried out near-infrared photometry for fainter objects
to probe the contamination at lower masses and across the
stellar/substellar boundary.

Near-infrared ($K'$-band) photometry was obtained for 29
optically-selected candidates on 10--12 July 2003
with the CFHTIR infrared camera.
This camera has
a 1024\,$\times$\,1024 pixel HAWAII detector with a spatial scale of
0.204$\arcsec$/pixel yielding a 3.5\,$\times$\,3.5 arcmin
field-of-view. The total exposure time was on the order of 5 minutes
per object and the photometric errors better than 0.1 mag.

Addtional infrared photometry ($K_{s}$-band) was obtained for
36 candidates on 10--13 June 2004 with MAGIC
on the Calar Alto 2.2-m telescope.
The MAGIC camera has a Rockwell 256\,$\times$\,256 pixel NICMOS3 array
with a spatial scale of 0.64$\arcsec$/pixel yielding a 164$\arcsec$
field-of-view in the high-resolution mode when mounted on the
2.2-m. Five dithered frames offset by about 25 pixels were
taken for each target, yielding exposure times between 20 and 120
seconds depending on the brightness of the target.

The infrared data was reduced following standard procedures.
First, a sky image was created using a median of the dithered
frames. Then, each science frame was sky-subtracted and flat-fielded.
The flat-field is the difference between averaged dome flat fields
observed lamp on and lamp off.
The measured magnitudes were corrected for extinction
and exposure time.
The airmass correction was assumed to be 0.07 and 0.088 mag/airmass
for Mauna Kea and Calar Alto, respectively.
Zero-points from the various standards observed throughout
the nights were applied to the instrumental magnitudes and
cross-checked with 2MASS to derive the final magnitudes.

\section{Contamination of the optical sample}
\label{cr359:IRfollowup_contamination}
\subsection{Near-infrared photometry}

\begin{figure}
\begin{center}
\includegraphics[angle=0, width=\linewidth]{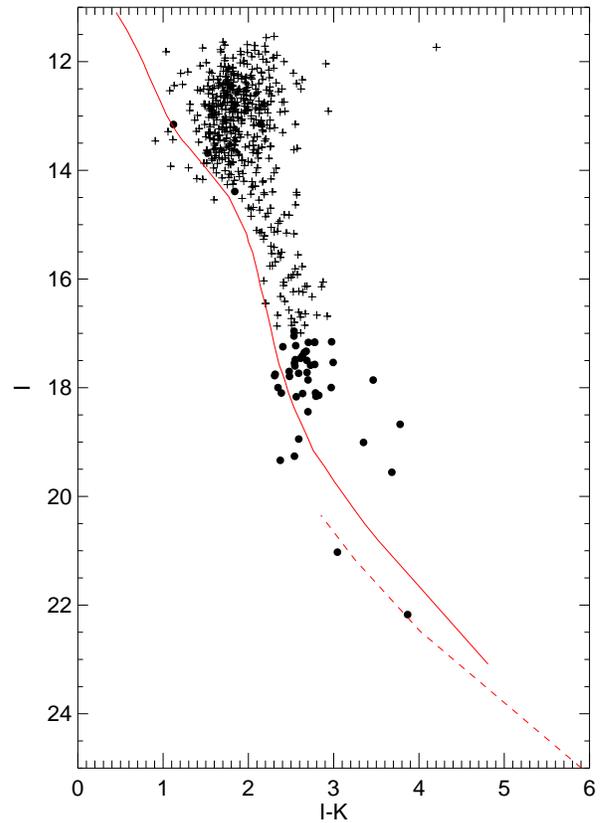}
\caption{
Colour-magnitude diagram ($I-K$,$I$)   
for the optically-selected candidates in
\Coll{}. The plus symbols denotes candidates with
2MASS photometry whereas the filled circles have been
observed with the CFHTIR and MAGIC cameras.
Overplotted are NextGen \citep[solid line;][]{baraffe98},
and the DUSTY \citep[dashed line;][]{chabrier00c}, assuming
an age of 60 Myr and a distance of 450 pc for \Coll{}.
}
\label{fig_cr359:IIKcand_CMD}
\end{center}
\end{figure}

Figure \ref{fig_cr359:IIKcand_CMD} shows the location of the optically-selected
candidates in \Coll{} in an optical-to-infrared \CMD{} ($I-K$,$I$).
Objects with 2MASS magnitudes are displayed with a plus sign while
targets from our own near-infrared follow-up observations
are shown as filled circles. The solid and dashed lines represent
the 60 Myr NextGen and DUSTY isochrones, respectively, at a distance
of 450 pc.
The isochrones are drawn using the $I$ filter from the CFH12K camera and 
the $K_{s}$ filter from 2MASS (I.\ Baraffe, personal communication).
Note that the $K'$ and $K_{s}$ magnitudes are very similar and
do not affect the results concerning the membership of the
candidates.

According to their location in the ($I-K$,$I$) diagram,
cluster candidates are divided into two subsamples as follows:
\begin{enumerate}
\item Probable members (Y$+$): these objects lie to the right of the
NextGen$+$DUSTY isochrones, shifted at a distance of 450 pc.
Their colours are consistent with cluster membership.
\item Non-members (NM): objects with colours bluer than those predicted
by the isochrones shifted at a distance of 450 pc.
\end{enumerate}

%
%

Combining the optical and optical-to-infrared \CMDs{},
most of the candidates brighter than $I$ = 17.0 mag
remain possible candidates. Indeed, only 9 out of 433
are rejected as cluster members, yielding a contamination
on the order of few percent. This simple analysis is however inadequate.
Both colour-magnitude diagrams show a wide cluster sequence at
$I$\,$\leq$\,14 mag, suggesting a large contamination by
red field stars. Using our current data we are unable to estimate
the contamination properly in this magnitude range.
Ultimately, low-resolution optical spectroscopy will provide
additional constraints, including lithium, radial velocities, and 
spectral types, to distinguish field stars from cluster members.

At fainter magnitudes, the cluster sequence becomes clearer and
extends down to $I$ = 22 mag in the optical \CMD{}. 
From the optical-to-infrared \CMD{},
we rejected 6 optically-selected cluster candidates out of 42 in the
magnitude range $I$ = 17--22 mag, yielding a
contamination of $\sim$\,14\,\% across the stellar/substellar boundary.
It is hard to set limits on the contamination at present
due to the small number statistics across the stellar/substellar
boundary. We are still lacking photometry for cluster 
candidates fainter than $I$ = 17 mag.

\subsection{Statistical contamination from a control field}

We now estimate the statistical contamination from a control field
observed within the framework of the CFH12K optical survey conducted in IC\,4665
\citep{deWit05}. The control field coordinates are
R.A.\ = 17$^{\rm h}$40$^{\rm m}$53$^{\rm s}$ and dec = $+$02$^{\circ}$50$'$14$''$
corresponding to a Galactic latitude of $b$ = $+$17$^{\circ}$ similar
to IC\,4665 but higher than \Coll{} ($b$ = $+$12.5$^{\circ}$).
The study of IC\,4665 reveals a contamination of
85\,\% and 70\,\% in the low-mass and brown dwarf regimes, respectively
\citep[see][for more details]{deWit05}. Those figures are likely
lower limits for \Coll{} due to its lower Galactic latitude and larger
distance compared to IC\,4665 (450 pc vs.\ 350 pc).

%
%
\section{IC\,4665 \& Collinder 359: a larger picture}
\label{cr359:ic4665_cr359}

\begin{figure}[!h]
\begin{center}
\includegraphics[angle=0, width=\linewidth]{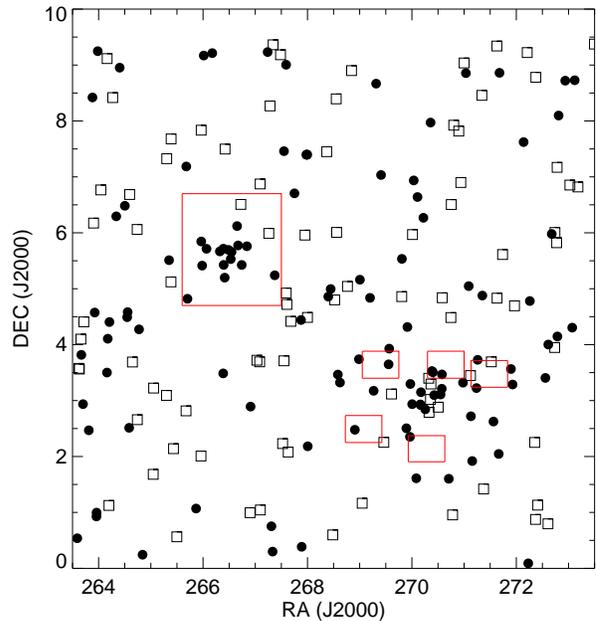}
\caption{
Sources brighter than $V$ = 10.5 mag with proper motion from
the ASCC2.5 \citep{kharchenko01} 
over a 100 square degree area encompassing IC\,4665 and \Coll{}.
Filled circles correspond to sources with a proper motion
consistent with the cluster motions (0.0,$-$9.0) mas/yr.
Open squares are sources with a random motion.
The filled circles are grouped into two clusterings whereas
open squares are randomly distributed over the 10 by 10 degree 
area. The black boxes correspond to our CFH12K coverage of the open clusters
\Coll{} and IC\,4665\@.
}
\label{fig_cr359:large_ra_dec}
\end{center}
\end{figure}

The open clusters \Coll{} and IC\,4665 are very close on the sky, their
centres being about 5 degrees apart (Fig.\ \ref{fig_cr359:large_ra_dec}).
Additionally, they share the
same proper motion \citep[0.0,$-$9.0 mas/yr;][]{perryman97}, have
similar ages (50--100 Myr) and comparable distance estimates
within the uncertainty errors (450 vs.\ 350 pc).

We have selected all sources brighter than
$V$ = 10.5 mag in a 10$\times$10 degrees
area encompassing \Coll{} and IC\,4665 using archival data from the
ASCC2.5 catalogue \citep{kharchenko01}. This catalogue is a compilation
of high-precision catalogues from space missions, including HIPPARCOS and Tycho-2,
and ground-based proper motions surveys providing coordinates,
accurate proper motions, and photometry. Sources located with a 2.5 mas/yr
radius and centred on (0.0,$-$9.0) mas/yr in right ascension
and declination, respectively, define two
clusterings (filled circles in Fig.\ \ref{fig_cr359:large_ra_dec})
at the nominal cluster centres.
In comparison, objects with proper motion of $-$3.0 and 0.0 mas/yr
in right ascension and declination, respectively, are randomly distributed
over the same area (squares in Fig.\ \ref{fig_cr359:large_ra_dec}),
implying that we have indeed two clusters.

%
%
%

%
%
\begin{figure}[!h]
\begin{center}
\includegraphics[angle=0, width=\linewidth]{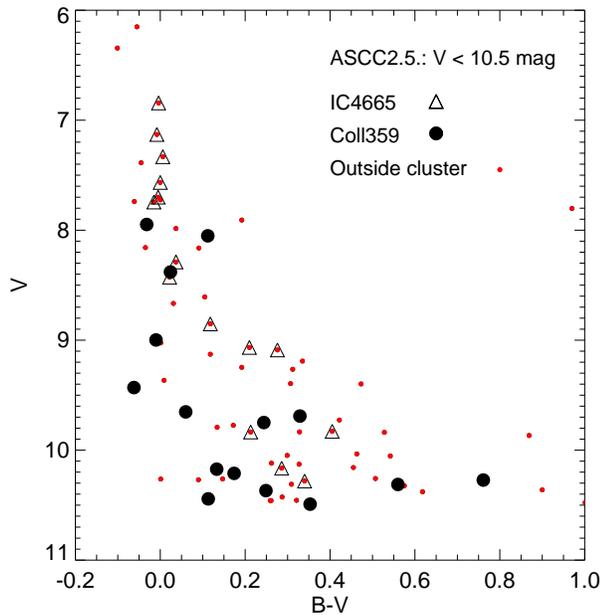}
\caption{
Colour-magnitude diagram ($B-V$,$V$) for sources located within
a two degrees box around the cluster centres and with a
proper motion consistent with IC\,4665 (open triangles) 
and \Coll{} (filled circles).
The small circles represent the objects with proper motion
similar to the two open clusters but outside the two degree
boxes around each cluster centre.
}
\label{fig_cr359:BVV_PM_CMD}
\end{center}
\end{figure}

We have examined the number of objects located within a 2 degree box
around each cluster. We have found 15 and 19 sources 
(filled circles in Fig.\ \ref{fig_cr359:large_ra_dec}) in IC\,4665 and
\Coll{}, respectively, and a total of 2 and 7 field dwarfs 
(open squares in Fig.\ \ref{fig_cr359:large_ra_dec}) in the
same area around each cluster. From the photometry point of view,
15 and 16 sources have $B-V$\,$<$\,1.0 in IC\,4665 and \Coll{}, respectively.
The number of field dwarfs with $B-V$\,$<$\,1.0 falls down to 1 and 4, respectively.
Moreover, the objects in \Coll{}
do not defined a sequence in the ($B-V$,$V$) colour-magnitude
diagram as clear as the members of IC\,4665 do (Fig.\ \ref{fig_cr359:BVV_PM_CMD}).
The lack of clear sequence in the colour-magnitude diagram and
the larger number of contaminants towards the line of sight of
\Coll{} tend to suggest that this cluster is much less massive than
IC\,4665\@. Nevertheless, the shift present between both sequences
could well be assigned to an extinction of $\sim$0.1 mag as the 
region exhibit signs of variable extinction.
Moreover, \Coll{} might not be a real cluster but either
linked to IC\,4665 or a moving group associated to the Ophiuchus
cloud \citep{roeser94}. A survey of the inner region 
of the cluster i.e.\ within 1 degree
radius of the cluster centre is required to answer those issues
and determine the characteristics of \Coll{}.

%
%
\section{Summary}
\label{cr359:summary}

We have presented the
first deep optical wide-field imaging survey complemented
with near-infrared photometric observations of the young open
cluster \Coll{}. We have surveyed a 1.6 square degree area in the
cluster in the $I$ and $z$ filters down to detection
and completeness limits of 22.0 and 24.0 mag with the CFH12K wide-field camera
on the Canada-France-Hawaii 3.6-m telescope. Based on their location 
in the optical ($I-z$,$I$) \CMD{}, we have extracted a total of 506
cluster member candidates
in \Coll{} spanning 1.3--0.030\,\Msun{}, assuming a distance
of 450 pc and an age of 60 Myr. The uncertainties on the distance
and age are 200 pc and 20 Myr, respectively.
We have cross-correlated the optically-selected candidates with
the 2MASS database for objects brighter than $I$ = 17.0 mag to weed
out a proportion of contaminating field stars. 
Further $K'$-band photometry has been
obtained for a subsample of 49 faint cluster candidates to
probe the contamination at and below the stellar/substellar
boundary.

By comparing the location of the brightest
cluster member 67 Oph, with solar metallicity isochrones
including moderate overshoot, we have
derived an age of 60\,$\pm$\,20 Myr for \Coll{}.
Taking into account the observed differences in ages of open clusters
between the turn-off main-sequence method and the lithium test,
\Coll{} is likely older than \APer{} but younger than the Pleiades.
Thus, the expected age of the cluster is at least twice larger 
than previously thought.
Finally, the comparison of the number of sources in a control
field with the number of selected cluster candidates indicates
that the surveyed fields do not contribute significantly to
the cluster population. A new optical survey closer to the cluster
centre but avoiding bright stars is needed to assess its extent
and properties.

%
%
\begin{acknowledgements}
This CFHT Key Programme (Bouvier, PI) was conducted within the framework of
 the European Research Training Network (McCaughrean, coordinator)
entitled ``The Formation and Evolution of Young Stellar Clusters''
(HPRN-CT-2000-00155) which we acknowledge for support. 
NL acknowledge funding from PPARC UK
in the form of a research associate postdoctoral position.
NL thank Nina Kharchenko and Anatoly Piskunov for useful discussion
on Collinder 359 during their stay at the AIP in Potsdam in spring 2004\@.
We are grateful to Isabelle Baraffe for providing us with the
NextGen and DUSTY models for the CFHT filters and Emmanuel Bertin
for supplying the PSFex package.
The authors wish to extend special thanks to those of Hawaiian 
ancestry on whose sacred mountain we are privileged to be guests.
This research has made use of the Simbad database, operated at
the Centre de Donn\'ees Astronomiques de Strasbourg (CDS), and
of NASA's Astrophysics Data System Bibliographic Services (ADS).
This research
has also made use of data products from the Two Micron All Sky Survey, which
is a joint project of the University of Massachusetts and the Infrared
Processing and Analysis Center, funded by the National Aeronautics and Space 
Administration and the National Science Foundation.
\end{acknowledgements}

%
%
 \bibliographystyle{aa}
 \bibliography{/local/data/68/nl41/publications/AA/mnemonic,/local/data/68/nl41/publications/AA/biblio_old}

\end{document}